\documentclass[12pt]{article}
\usepackage{sc3conf}
\usepackage{amsfonts}
\usepackage{epsfig}

\begin{document}

\title{Lightcurves of thermonuclear supernovae as a probe of the
explosion mechanism and their use in cosmology}
\authors{S.I.~Blinnikov\adref{1}
  and E.I.~Sorokina\adref{2}}
\addresses{\1ad ITEP, 117218, Moscow, Russia,
\nextaddress \2ad Sternberg Astronomical Institute, 119992 Moscow, Russia}

\maketitle              

\begin{abstract}
 Thermonuclear supernovae are valuable for cosmology
but their physics is not yet fully understood.
Modeling the development and propagation
of nuclear flame is complicated by numerous instabilities.
The predictions of supernova light curves still involve
some simplifying assumptions, but one can use the comparison
of the computed fluxes with observations to constrain the explosion
mechanism.
In spite of great progress in recent years, a number of issues
remains unsolved both in flame physics and light curve modeling.

\end{abstract}

\section{Introduction}

Supernovae of type Ia (SNe~Ia) are important for cosmology (better to say,
for cosmography) due to their brightness.
They are not standard candles, but can be used for measuring distances
with the help of the peak luminosity -- decline rate
correlation, established by Yu.P.~Pskovskii \cite{psk77}
and M.M.~Phillips \cite{phil93}
(see the review \cite{leib00}).
To exclude systematic effects in linking the observed light of distant SNe~Ia
to the parameters of cosmological models, one has to understand the nature
of supernova outbursts and to build accurate algorithms for predicting their
emission.

This involves:
 identifying the progenitors of SNe~Ia;
 the birth of
 thermonuclear flame and its accelerated propagation leading to explosion;
 light curve and spectra modeling.

When the full understanding
will be achieved, one can try to evaluate
the importance of evolution effects in using supernovae as distance
indicators.
In spite of great progress in recent years, a number of issues
remains unsolved both in flame physics and light curve modeling.
We point out some problems which seem most important to us.

\section{Progenitors}

Mechanical
equilibrium and evolution of stars is easily understood from
the virial theorem for a star:
$
  3 \int \! PdV = -U \; ,
$
where $P$ is the pressure, $V$ is volume, and $U$
is  the gravitational energy of a star.
Crude estimates
$ V \sim R^3, \quad
  M \sim R^3\rho, \quad U \sim - GM^2/R
$ in the virial equation give
$
P  \sim GM^{2/3}\rho^{4/3}\;.
$
For an ideal classical gas, the equation of state
$P=\mathcal{R}\rho T$ implies
$
 \mathcal{R}T \sim GM^{2/3}\rho^{1/3},
$
with $\mathcal{R}$ the gas constant.
This is already enough to understand the evolution of massive stars! 
While a star loses energy and contracts, its internal temperature $T$ grows.
If the losses are balanced by a nuclear energy release, then the contraction stops
and a thermal equilibrium is established:
$$
\mathrm{nuclear \; heating \; power}\; {L^+} =  \mathrm{radiative \; cooling \; (luminosity)} \;
{L^-}.
$$
The rate of thermonuclear heating scales as
$\langle \sigma v_0\rangle \sim \exp[-(\alpha_G/T)^{1/3}]$
due to the Gamow's peak:
the chances to penetrate the Coulomb barrier for fast nuclei
grow, but the tail of Maxwell distribution goes down.
Here $\alpha_G$ depends strongly on nuclei charges $Z_i$:
$\alpha_G \propto Z_1^2Z_2^2 $, 
thus high-$Z$ ions can fuse only at high $T$. 
Small perturbations of $T$ produce huge variations
in {$L^+$} since, normally, { $T \ll \alpha_G$}.

The cooling $L^-$  depends on $T$ moderately, and it seems that $L^-$ cannot compensate
an overheating perturbation.
Why then do not all stars explode violently?
The reason is the same as for the growth of $T$ in stars losing the energy at
contraction: they have negative heat capacity.
For a star made of a classical plasma with $\gamma=5/3$ the internal
energy $Q= \frac{3}{2}{\mathcal R} M{T}$ and the virial theorem implies
$ U=-2Q $, so the total energy $\mathcal E$ is negative:
$
 {\mathcal E}=U+Q=-Q<0
$.
Thus any growth in  $\mathcal E$ due to heating leads to the drop of $T$ (because nuclear energy
is used for expansion against gravity).
The perturbations decay, and there is no hope to get a thermonuclear supernova
from a normal star composed of a classical plasma.

The situation changes, if a star is made of
a degenerate matter. Then
the heat $\frac{3}{2}\mathcal{R} M{T}$ resides in
ions, but its absolute value is much less than $Q$ which is now governed by Fermi energy
of electrons.
We have crudely for a mixture of non-relativistic electron Fermi gas and classical ions:
$
  P \sim K\rho^{5/3} + \mathcal{R}\rho T \sim GM^{2/3}\rho^{4/3},
$
and
$
  \mathcal{R}T \sim  GM^{2/3}\rho^{1/3} - K\rho^{2/3}.
$
So at high density, the equilibrium temperature $T$ decreases with growing $\rho$, i.e.
goes the same way as ${\mathcal E}$.
The total heat capacity becomes positive,
and runaway can set in as in terrestrial explosives.
So, a progenitor of SN~Ia must be a degenerate star - a white dwarf.

A single white dwarf is unable to explode, it cools down.
But when it is in a binary system the chances to produce a supernova do appear (we need
only one in $\sim 300$ dying white dwarfs to explode in order to explain the rate
of SNe~Ia).
Even if the binary has two dead white dwarfs, it can explode because they can
merge due to emission of gravitation waves (double-degenerate, or DD scenario \cite{iben97}).
If one star in the binary is alive, the white dwarf can accrete its lost mass
and reach an instability
(single-degenerate, or SD scenario \cite{whelib73,brag90}).
It is unclear which scenario is most important, there are strong arguments\cite{kob98}
from chemical evolution that only SD is the viable one.
On the other hand, it seems that DD can produce a richer variety of SN~Ia events.
Moreover,
discoveries of intergalactic SNe~Ia \cite{bart97,galgf02} can be explained more
naturally, because a DD system may evaporate from a galaxy.
It is quite likely that both scenarios are being played, but their relative role
may change in young and old galaxies. If so, a systematic trend may appear in SNe~Ia
properties with the age of Universe, and this may have important consequences for
cosmology.

\section{Thermonuclear flames}

After merging  in DD scenario, or after the white dwarf accretes large amount of material
in SD case, the explosive instability  develops.
In principle, combustion can
propagate either in
the form of a supersonic  \emph{detonation} \cite{arn69} wave, or
as a subsonic \emph{deflagration} \cite{iich74,nomsn76} (flame).
In detonation, the unburned fuel is
ignited  by a shock front propagating
ahead of the burning zone itself.
In {deflagration}, the ignition
is governed by heat and active reactant transport, i.e.
by thermal conduction and diffusion.

Most likely, the runaway starts as a laminar flame propagating due
to thermal conduction. In terrestrial flames,
the `fusion' of molecules goes with the rate: 
$
 \langle \sigma v_0\rangle \sim  \exp ( -{E_a / {\mathcal R} T}),
$
-- the Arrhenius law of chemical burning.
Here {$E_a$} is \emph{activation energy}.
The parameter, showing the strong $T$-dependence of the heating
$
 \mathrm{Ze}=
 {\partial \log \langle \sigma v_0\rangle   / \partial \log  T} \simeq
 { E_a / {\mathcal R} T}
$
is called the Zeldovich number in the theory of chemical flames.
For them typically $\mathrm{Ze} \sim 10 \dots 20$.
The classical theory \cite{zfk38}
predicts the flame speed
$
 v_\mathrm{f} \approx \mathrm{Ze}^{-1}[{v_T l_T}/ \tau_\mathrm{reac}(T_b)]^{1/2}
$,
with
$
 \tau_\mathrm{reac}(T) \propto \exp[ E_a /({\mathcal R} T)]
$.
In SNe, for nuclear flames,
$
 \tau_\mathrm{reac}(T) \propto \exp[\alpha_G^{1/3} /( 3 T^{1/3})]
$,
and,
$
 \mathrm{Ze}=
 {\partial \log \langle \sigma v_0\rangle / \partial \log  T} \simeq
  \alpha_G^{1/3} /(3 T^{1/3})
$,
which has values very similar to terrestrial chemical flames.

A big difference with chemical flames is the ratio of heat
conduction and mass diffusion,
the Lewis number,
$
  \mathrm{Le}= (v_T l_T) / (v_D l_D)
$.
One finds
$\mathrm{Le}\sim 1$ in laboratory gaseous flames, while
$\mathrm{Le}\sim 10^7$ in thermonuclear SNe, since
heat is transported by relativistic electrons,
$v_T \sim c$, and there is almost no diffusion, $l_T \gg  l_D$.
Nevertheless, the modern computations \cite{timw92} follow
the old theory \cite{zfk38} closely.
The conductive flame propagates in a presupernova with  $v_{\rm f}$
which is too slow to produce an energetic explosion:
the ratio of $v_{\rm f}$  to sound
speed, i.e. the Mach number, Ma, is very small (see Table 1). 
The star has enough time to expand, to cool down, and
the burning dies completely.
So an acceleration of the flame is necessary in order to explain the
SN phenomenon. This is the main problem in current research of SNe~Ia hydrodynamics.

\begin{table}
\begin{center}
\caption{Flame speed $v_{\mathrm f}$ and width $l_{\mathrm f}$
 in C+O \protect\cite{timw92}}
\begin{tabular}{|lllll|}
\hline\noalign{\smallskip}
 $\rho $ & $v_{\mathrm f}$ & $l_{\mathrm f}$ & $ \Delta\rho/\rho$   & Ma  \\
 $10^9$ gcc  & km/s & cm & &   \\
\noalign{\smallskip}
\hline
\noalign{\smallskip}
6  &  214 &  $ 1.8\times 10^{-5}$  &  0.10  &  $2\times 10^{-2}$ \\
1  &  36 &  $ 2.9\times 10^{-4}$  &  0.19  &  $4\times 10^{-3}$ \\
0.1  &  2.3 &  $ 2.7\times 10^{-2}$  &  0.43  &  $4\times 10^{-4}$ \\
\hline
\noalign{\smallskip}
\end{tabular}
\end{center}
\label{Tab1}
\end{table}


There is a rich variety of instabilities that can severely distort the shape
of a laminar flame.
The Rayleigh--Taylor
(RT) instability governs the corrugation of the front on
the largest scales. On the smallest scales the flame is
controlled by the  Landau-Darrieus (LD) instability.
RT, LD instabilities  and turbulence make computations difficult,
but without them a star would not explode. All these  instabilities were
considered  already by L.Landau \cite{lan44} as a means to accelerate the
flame.

Because of instabilities, the flame surface  becomes
wrinkled and its area
grows as
$
  S \propto \bar R^\alpha \; ,
$
with average radius $\bar R$ and $\alpha > 2$, i.e.
faster than $S \propto \bar R^2$. 
In other words the surface  becomes
`fractal'. The exponent {$\alpha$}
is actually the fractal dimension, {$\alpha = D_{\rm F}$}.
The effective flame speed is determined \cite{woo90}  
by the ratio of the maximum scale of the instability to the minimum one:
$
 v_{\rm eff}= v_{\rm f}(\lambda_{\rm max}/\lambda_{\rm min})^{D_{\rm F}-2}.
$
When the vorticity is not important  it is possible to study in detail the
non-linear stage of LD instability and to find the fractal
dimension \cite{BS96}. A similar dependence of the flame fractal dimension on
the density jump across the front was found in SPH simulations of the flame
subject to RT instability \cite{BrGar95}.

The fractal description is good for LD while it remains mild, because it
operates in a star on the scales from the flame thickness (a tiny fraction of
a cm) up to $\sim 1$ km. For the RT instability,
{$\lambda_{\rm max}/\lambda_{\rm min}$} is very uncertain
and the fractal dimension is uncertain too. So a direct 3D numerical
simulation is necessary. The same is true for a low density regime of LD when
it is strongly coupled to turbulence (generated on the front, or cascading
from large RT vortices). A great progress is achieved here in several
groups \cite{HN00,martin,fritz,kho00}.
When simulating 3D turbulent
deflagrations one encounters two problems: the
representation of the thin moving surface separating hot and
cold material, and the prescription of the
local velocity $v_\mathrm{f}$ of this surface as a
function of the large-scale flow
with a crude numerical resolution $>1$ km. One solution  is
sketched in \cite{martin}; for a different approach see \cite{kho00}.
In spite of the progress this problem cannot be treated as completely solved,
and even 1D approach may give
interesting results, especially for unusual SNe~Ia \cite{dunib01}.

\section{Light curves of SNe~Ia}

Given a hydrodynamic structure of SN ejecta, one can compute
a light curve which should be compared
with observations.
There are several effects in SNe physics which lead to difficulties
in the light curve modeling of any type of SNe.
For instance, an account should be taken correctly for deposition
of gamma photons produced in decays of radioactive isotopes, mostly $^{56}$Ni
and  $^{56}$Co.
To find this one has to solve
the transfer equation for gamma photons together with hydrodynamical
equations.
Full system of equations should involve also radiative transfer equations
in the range from soft X-rays to infrared for the expanding medium.
There are millions of spectral lines that form SN spectra, and it is
not a trivial problem to find a convenient way how to treat them
even in the static case.
The expansion makes the problem much more difficult to solve:
hundreds or even thousands of lines give their input into emission and
absorption at each frequency.

Currently, powerful codes appear aimed to attack
a full 3D time-dependent problem of SN~Ia light \cite{hoeflich}.
Yet there are some basic questions, like averaging the line opacity
in expanding media, that remain controversial.

In our work we predict the broad-band UBVI and bolometric light curves of SNe
Ia, using our 1D-hydro code which models multi-group time-dependent
non-equilibrium radiative transfer inside SN ejecta.  In our previous analysis
we have studied two Chandrasekhar-mass models: the classical deflagration
model W7  \cite{W7} and the delayed detonation one DD4 \cite{DD4}, as well as
two sub-Chandrasekhar-mass models: helium detonation model LA4 \cite{livne}
and low-mass detonation model  WD065 with low $^{56}$Ni production
\cite{pilar}, which was constructed for modelling subluminous SNe~Ia events,
such as SN~1991bg. All those models were simplified spherically-symmetrical
(1D) ones.

\begin{figure}[ht]
  \centerline{\includegraphics[width=0.8\textwidth]{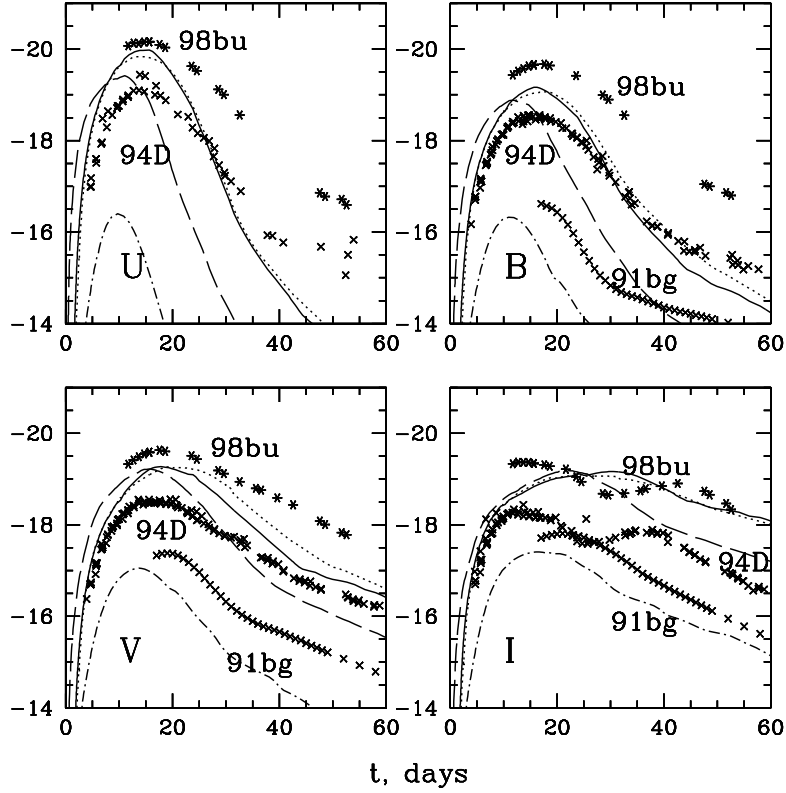}}
  \caption{UBVI light curves for 4 1D models (W7 -- solid line, DD4 --dots,
LA4 -- dashes, WD065 -- dash-dots).
  Crosses, stars and triangles show the light curves
for three observed SNe~Ia.}
  \label{fig1DLCs}
\end{figure}

The UBVI light curves of 1D models are shown in Fig.~\ref{fig1DLCs}.
The Chandrasekhar-mass models demonstrate almost identical light curves.
The sub-Chandrasekhar-mass ones are more different.
WD065 has almost similar element distribution as Chandrasekhar-mass
models, and the shape of its light curve is in principle the same as that of
W7 and DD4.
It is just much dimmer due to a very low $^{56}$Ni
abundance.

LA4 is very different from any other model, since the explosion there
started on the surface of a white dwarf, not in the center,
as for every other model, so there is a $^{56}$Ni layer near the surface
in LA4.
This feature explains why the model is essentially bluer
than  other ones.

\begin{table}
\begin{center}
\caption{Parameters of SN Ia models}
\label{models}
\begin{tabular}{llllll}
\hline
Model & DD4 & W7 & LA4 & WD065 & MR \\
\hline
$M_{\rm WD}{}^{\rm a}$      & 1.3861 & 1.3775 & 0.8678 & 0.6500  & 1.4 \\
$M_{{}^{56}{\rm Ni}}{}^{\rm a}$ & 0.63 & 0.60 & 0.47   & 0.05  & 0.42 \\
$E_{51}{}^{\rm b}$ & 1.23   & 1.20   & 1.15   & 0.56  & 0.46\\
\hline
\multicolumn{5}{l}{${}^{\rm a}$in $M_\odot$} \\
\multicolumn{5}{l}{${}^{\rm b}$in $10^{51}$~ergs~s${}^{-1}$}
\end{tabular}
\end{center}
\end{table}

Currently we employ our new corrected treatment for line
opacity \cite{SBring02} in the expanding medium, which is important
especially in UV and IR bands.
It seems that 1D thermonuclear supernova models, e.g.
the deflagration W7 \cite{W7} model and the delayed detonation DD4 \cite{DD4}
one, produce the light curves fitting the observations not so good as the
recent 3D deflagration model MR computed at MPA \cite{martin}. 
We believe that the main feature of the latter model
which allows us to get the correct flux during the first month, is strong
mixing that moves the material enriched with radioactive
$^{56}$Ni to the outermost layers of SN ejecta.

\begin{figure}[ht]
  \centerline{\includegraphics[width=0.8\textwidth]{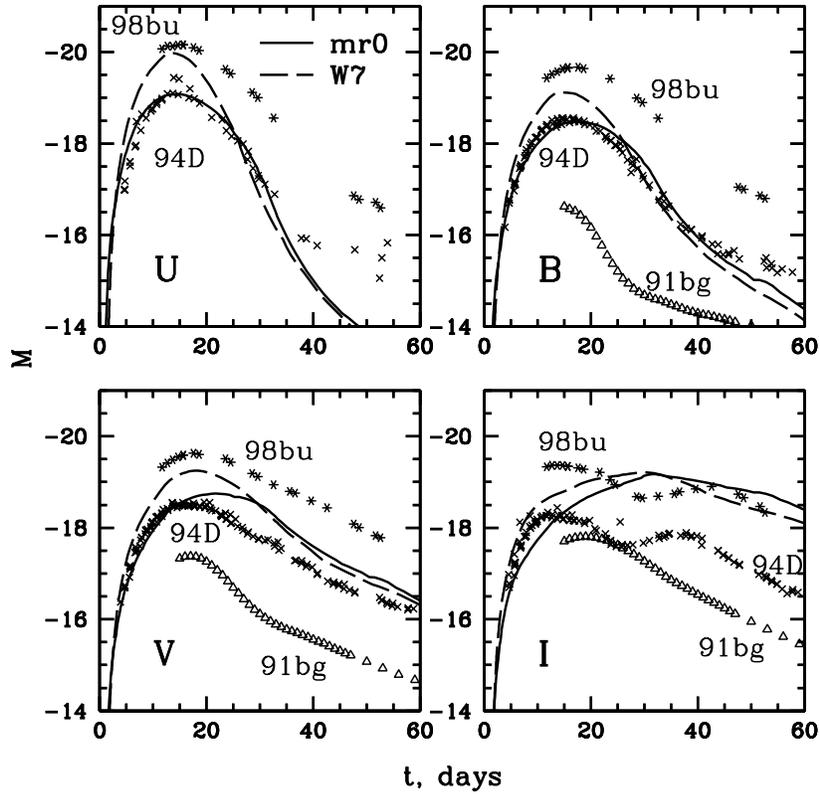} 
}
  \caption{UBVI light curves for the 3D (MR; solid) and 1D (W7; dashed) models.
  Crosses, stars and triangles show the light curves
for three observed SNe~Ia.}
  \label{figw7mr0}
\end{figure}

Fig.~\ref{figw7mr0}
demonstrates that in spite of quite different
structure of the old W7 model and the new MR one their light curves
are similar in many details.
Moreover the new model behaves better in $U$ and $B$ bands.
Unfortunately, the bolometric light curve for MR model
is somewhat too slow.
The ejecta must expand with a higher speed to let photons to diffuse out
faster.

\section{Conclusions}

There are several points which require attention for applying SNe~Ia
in cosmology: progenitors may be different in younger galaxies;
burning regimes may change with the age of Universe \cite{ourLC}.
The physical understanding of the Pskovskii-Phillips is not yet achieved.

In the flame modeling the new 3D SN~Ia model \cite{martin} is very appealing.
Yet it is not a final one: a detailed post-processing of nucleosynthesis
is not yet checked in the light curve calculation.

The SN light curve modeling still has a  lot of physics to be added, such as a
3D time-dependent radiative transfer, including as much as possible of NLTE
effects \cite{hoeflich}, which are especially essential for SNe~Ia. All this
will improve our understanding of thermonuclear supernovae and their role in
cosmology.

Our work is partly funded by RFBR (grants 00-02-17230 and 02-02-16500).
SB is grateful to Ana Mourao for support.


\begin{thebibliography}{99}
\addcontentsline{toc}{section}{References}


\bibitem{arn69} W.D.~Arnett, {\it Ap.Sp.Sci.}  \textbf{5}, 180 (1969).
\bibitem{bart97} O.~Bartunov, Outlying Supernovae - Myth or Reality?
{\it UCSB Workshop on SNe}, {\tt http://www.sai.msu.su/$\sim$megera/sn/outsn/} (1997).
\bibitem{BS96} S.I.~Blinnikov, P.V.~Sasorov, {\it Phys.Rev.} \textbf{E53} 4827  (1996).
\bibitem{BrGar95} E.~Bravo, D.~Garc\'ia-Senz, {\it ApJ} \textbf{450}, L17 (1995).
\bibitem{brag90} A.~Bragaglia et al.,
{\it ApJLett} \textbf{365}, L13 (1990).
\bibitem{dunib01} N.V.~Dunina-Barkovskaya et al., 
{\it Astron.Letters}  \textbf{27}, 353 (2001).
\bibitem{galgf02}
A.~Gal-Yam et al., 
astro-ph/0211334 (2002).
\bibitem{leib00} B.~Leibundgut 
{\it Astr.Ap.Rev.} \textbf{10}, 179 (2000).
\bibitem{HN00}
W.~Hillebrandt, J.~C.Niemeyer,  {\it Ann.~Rev.~Astron.~Ap.}
  \textbf{38}, 191 (2000).
\bibitem{hoeflich}
P.~H{\" o}flich,
Workshop on Stellar Atmosphere
Modeling,  Eds: I.~Hubeny et al., astro-ph/0207103 (2002).
\bibitem{iben97} I.J.~Iben,  et al.,
{\it ApJ} \textbf{475}, 291  (1997).
\bibitem{iich74} L.N.~Ivanova et al., 
     {\it Space Sci.}  \textbf{31}, 497 (1974).
\bibitem{kho00} A.~Khokhlov, e-print astro-ph/0008463 (2000).
\bibitem{kob98} C.~Kobayashi et al.,
{\it ApJLett} \textbf{503}, L155 (1998).
\bibitem{lan44} L.D.~Landau, 
  {\it Acta Physicochim. USSR} \textbf{19}, 77  (1944)
\bibitem{livne} E.~Livne, D.~Arnett: {\it ApJ} \textbf{452}, 62
  (1995)
\bibitem{nomsn76} K.~Nomoto et al., 
    {\it Ap.Space Sci.}  \textbf{39}, L37 (1976).
\bibitem{W7} K.~Nomoto et al., 
 {\it ApJ} \textbf{286}, 644 (1984).
\bibitem{phil93} M.M.~Phillips, {\it ApJ},  \textbf{413}, L105 (1993).
\bibitem{pilar} P.~Ruiz-Lapuente et al.: {\it Nature} \textbf{365},728 (1993)
\bibitem{psk77} Yu.P.~Pskovskii, {\it Sov.Astronomy} \textbf{21}, 675 (1977).
\bibitem{martin} M.~Reinecke et al.
{\it Astr.Ap.} \textbf{386}, 936 (2002).
\bibitem{fritz}
F.K.~R\"{o}pke, et al. 
{\it Proc. 11th Workshop Nuclear Astrophysics}, 
2002, W.Hillebrandt and E. M\"{u}ller (Eds.).
MPA/P13 (2002), p. 41 (astro-ph/0204036).
\bibitem{SBring02} E.I.~Sorokina, S.I.~Blinnikov: 
{\it Proc. 11th Workshop Nuclear Astrophysics}, 
2002, W.Hillebrandt and E. M\"{u}ller (Eds.).
MPA/P13 (2002), p.57 (astro-ph/0212187).
\bibitem{ourLC} E.I.~Sorokina et al., 
  {\it Astron. Letters} \textbf{26}, 67 (2000)
\bibitem{timw92}
F.X.~Timmes, S.E.~Woosley, {\it ApJ} \textbf{396}, 649 (1992).
\bibitem{whelib73} J.~Whelan,
I.J.~Iben 1973.
ApJ \textbf{186}, 1007-1014.
\bibitem{woo90} S.E.~Woosley,  in: {\it Supernovae,} ed.\ A.~G.~Petschek,
                 A \& A library, 1990, p.\ 182
\bibitem{DD4} S.E.~Woosley, T.A.~Weaver. In: \emph{Supernovae},
  ed. by J.~Audouze et al., Elsevier Science Publishers, Amsterdam (1994), p.63
\bibitem{zfk38} Ya.B.~Zeldovich, D.A.~Frank-Kamenetsky,
 {\it Acta Physicochim. USSR}  \textbf{9}, 341 (1938)

\end{thebibliography}
\end{document}